\documentclass[11pt,a4paper,fleqn]{article}

\usepackage{amsmath,amssymb,amsthm,enumerate,cite}

\newcommand{\vspacebefore}{\raisebox{0ex}[2.5ex][0ex]{\null}}

\newcommand{\p}{\partial}
\newcommand{\vk}{\varkappa}
\newcommand{\const}{\mathop{\rm const}\nolimits}
\newcommand{\Equiv}{\mathop{\rm \, equiv}}
\newcommand{\sign}{\mathop{\rm sign}\nolimits}
\newcommand{\thetbn}{\arabic{nomer}}

\newtheorem{theorem}{Theorem}
\newtheorem{notion}{Notion}

\newtheorem*{note*}{Note}
\newtheorem{lemma}{Lemma}

\begin{document}
\begin{center}
{\LARGE\bf New Results on Group Classification \\
of Nonlinear Diffusion-Convection equations\\[1.3ex]}
{\large\bf Roman~O.~Popovych~$^\dag$ and Nataliya~M.~Ivanova~$^\ddag$
\\[1.7ex]}
\footnotesize
Institute of Mathematics of National Academy of Sciences of Ukraine, \\
3, Tereshchenkivska Str., Kyiv-4, 01601, Ukraine\\
\vspace{1em}
E-mail: $^\dag$rop@imath.kiev.ua, $^\ddag$ivanova@imath.kiev.ua
\end{center}

\begin{abstract}
Using a new method and additional (conditional and partial) equivalence transformations,
we performed group classification in a class of variable coefficient
$(1+1)$-dimensional nonlinear diffusion-con\-vec\-tion equations of the
general form $f(x)u_t=(D(u)u_x)_x+K(u)u_x.$
We obtain new interesting cases of such equations with the
density $f$ localized in space, which have non-trivial invariance algebra.
Exact solutions of these equations are constructed.
We also consider the problem of investigation of the possible local transformations for an arbitrary
pair of equations from the class under consideration,
i.e. of describing all the possible partial equivalence transformations in
this class.

\end{abstract}

\section{Introduction}

The problems of group classification and exhaustive solutions of such problems
are interesting not only from purely
mathematical point of view, but is also important for applications.
In physical models there often exist a priori requirements to symmetry
groups that follow from physical laws (in particular, from Galilei or relativistic theory).
Moreover, modelling differential equations could contain parameters or functions
which have been found experimentally and so are not strictly fixed.
(It is said that these parameters and functions are arbitrary elements.)
At the same time mathematical models have to be simple enough
to analyze and solve them. Solving the problems of group
classification makes possible to accept for the criterion of applicability  the
following statement. Modelling differential equations have to admit a group
with certain properties or the most extensive symmetry group from the possible ones.

In this paper we consider a class of variable coefficient nonlinear
diffusion-convection equations of the form
\begin{equation} \label{eq1}
f(x)u_t=(g(x)D(u)u_x)_x+K(u)u_x,
\end{equation}
where $f=f(x),$ $g=g(x),$ $D=D(u)$ and $K=K(u)$ are arbitrary smooth functions of their variables,
$f(x)g(x)D(u)\!\neq\! 0.$
The linear case of~(\ref{eq1}) ($D,K=\const$) was studied by S.~Lie~\cite{lie1881}
in his classification of linear second-order PDEs with two independent variables.
(See also modern treatment of this subject in~\cite{Ovsiannikov1}.)
That is why we assume below that $(D_u,K_u)\ne(0,0),$
i.e.~(\ref{eq1}) is a nonlinear equation.

Moreover, using the transformation
$\;\tilde t=t,\; \tilde x=\int \frac{dx}{g(x)},\; \tilde u=u,\;$
we can reduce equation~(\ref{eq1})  to
\[
\tilde f(\tilde x)\tilde u_{\tilde t}= (D(\tilde u)
\tilde u_{\tilde x})_{\tilde x} + K(\tilde u)\tilde u_{\tilde x},
\]
where $\tilde f(\tilde x)=g(x)f(x)$ and $\tilde g(\tilde x)=1.$
(Likewise any equation of form (\ref{eq1})
can be reduced to the same form with $\tilde f(\tilde x)=1.$)
That is why  without loss of generality we restrict ourselves to investigation of the equation
\begin{equation} \label{maineq}
f(x)u_t=\left(D(u)u_x \right)_x + K(u)u_x.
\end{equation}

Apart from their own theoretical interest, equations~(\ref{maineq})
are used to model a wide variety of phenomena in physics,
chemistry, mathematical biology etc.
For the case $f(x)=1$ equation~(\ref{maineq}) describes
vertical one-dimensional transport of water in homogeneous
non-deformable porous media. When $K(u)=0$ this equation
describes stationary motion of a boundary layer of fluid over a flat
plate, an vortex of incompressible fluid in porous medium for
polytropic relations of gas density and pressure. The outstanding
representative of the class of equations (\ref{maineq}) is the Burgers
equation that is the mathematical model of a large
number of physical phenomena.
(For more detail refer to \cite{Ames,Barenblatt1,Barenblatt2,Malfliet}.)

Investigation of the nonlinear heat equations by means of symmetry methods started
in 1959 with the Ovsiannikov's work~\cite{Ovsiannikov} where he studied
symmetries of the equation
\begin{equation}
\label{Oveq}
u_t=(f(u)u_x)_x.
\end{equation}
In 1987 I.Sh.~Akhatov, R.K.~Gazizov and N.Kh.~Ibragimov~\cite{Akhatov&Gazizov&Ibragimov}
classified the equations
\begin{equation} \label{Akheq}
u_t=G(u_x)u_{xx}.
\end{equation}
V.A.~Dorodnitsyn (1982, \cite{Dorodnitsyn}) performed group classification of the equation
\begin{equation} \label{Doreq}
u_t=(G(u)u_x)_x+g(u).
\end{equation}
A.~Oron, P.~Rosenau (1986, \cite{Oron&Rosenau}) and M.P.~Edwards (1994, \cite{Edwards})
presented the most extensive at the time list of symmetries of the equations
\begin{equation} \label{Oroneq}
u_t=(G(u)u_x)_x+f(u)u_x.
\end{equation}
The  results of \cite{Ovsiannikov,Dorodnitsyn,Oron&Rosenau}
were generalized by R.M.~Cherniha and M.I.~Serov (1998,
\cite{Cherniga&Serov})
who classified the nonlinear heat equation with convection term
\begin{equation} \label{Cherneq}
u_t=(G(u)u_x)_x+f(u)u_x+g(u).
\end{equation}

It should be noted that equations~{(\ref{eq1})--(\ref{Cherneq})} are particular
cases of the more general class of equations
\begin{equation}\label{Zheq}
u_t=F(t,x,u,u_x)u_{xx}+G(t,x,u,u_x).
\end{equation}
Group classification of~(\ref{Zheq}) is presented in
\cite{Zhdanov&Lahno1999,Basarab-Horwath&Lahno&Zhdanov2001,%
Lahno&Samoilenko2002-I,Lahno&Samoilenko2002-II,Lahno&Spichak&Stognii2002}.
However, since the equivalence group of~(\ref{Zheq})
is essentially wider than those for {(\ref{eq1})--(\ref{Cherneq})}
the results of \cite{Zhdanov&Lahno1999,Basarab-Horwath&Lahno&Zhdanov2001,%
Lahno&Samoilenko2002-I,Lahno&Samoilenko2002-II,Lahno&Spichak&Stognii2002}
cannot be directly used
to symmetry classification of equations {(\ref{eq1})--(\ref{Cherneq})}.
Nevertheless, these results are
useful to find additional equivalence transformations in the above classes.

Equations of form~(\ref{maineq}) have been also investigated
from other than classic Lie symmetry points of view.
For instance, potential symmetries of subclasses of~(\ref{maineq})
where e.g. either $f=1$ or $K=0$ were studied
by C.~Sophocleous~\cite{Sophocleous1996,Sophocleous2000,Sophocleous2003}.

Some symmetry properties of class~(\ref{eq1}) were considered in a recent paper~\cite{El_labany}.
However, it does not present correct and complete results on the subject.
A overwhelming majority of cases with nontrivial Lie symmetry were omitted, and
there are mistakes in the cases that were adduced.
Nevertheless the subject seemed very interesting, and we decided
to study Lie symmetries of class~(\ref{maineq}).

The ultimate goal of this paper is to present an example of exhaustive solution
of the group classification problem in quite a difficult case.
After giving a precise definition and a discussion of this problem in general case,
we performed the complete extended group classification, found additional
equivalence transformations and exact solutions of equations~(\ref{maineq}).
A lot of new interesting cases of extensions of maximal Lie
symmetry group were obtained for these equations.
For example, we determined the equations which have the density $f$ localized in the space of $x$
and are invariant with respect to $m$-dimensional ($2\le m\le4$) Lie symmetry algebras,
that allows construction of new exact non-stationary solutions for them.

Problems of general group classification, except for really trivial cases, are very difficult.
It can be illustrated by the multitude of papers where such general classification problem
is solved incorrectly or incompletely.
There are also many papers on ``preliminary group classification'' where authors list some cases
with new symmetry but do not claim that the general classification problem is solved completely.
For this reason finding an effective approach
to simplification is essentially equivalent to feasibility of solving the problem at all.
In this paper we develop and apply a simple and effective tool based on investigation of
specific compatibility of classifying conditions. First it was proposed in~\cite{Nikitin}
and then applied to solving a number of different group classification problems~\cite{Popovych,Boyko,Vasilenko}.
The other tool is systematically to use additional (conditional and partial) equivalence transformations
that allows us to put in order, verify and analyze the obtained results.

\looseness=-1
Our paper is organized as follows. First of all (Section~\ref{sec_dom}) we
describe the group classification method used here and introduce
the notions of conditional and partial equivalence.
Then (Sections~\ref{sec_roc}) we significantly enhance the results of~\cite{El_labany}
and give the complete group classification of class~(\ref{maineq}).
Since the case $f(x)=1$ has a great variety of applications and was investigated earlier by a number of
authors, we collect results for this class together in Section~\ref{sec_f1}.
Section~\ref{sec_proof} contains the proof of the main theorem on group classification of the class~(\ref{maineq}).
We attempted to present our calculations in most reasonable detail so as it would be feasible to verify that.
Conditional equivalence transformations are considered in Section~\ref{sec.cets},
where we also present four lemmas on possible local equivalence transformations between two arbitrary
equations of form~(\ref{maineq}).
The results of group classification are used to find exact solutions of equations
from class~(\ref{maineq}) (Section~\ref{sec_exsol}).

\section{Group classification method and additional equivalence} \label{sec_dom}

Let us describe the classical algorithm of group classification restricting,
for simplicity, to the case of one differential equation of the form
\begin{equation}
\label{exameq}
L^\theta(x,u_{(n)})=L(x,u_{(n)},\theta(x,u_{(n)}))=0.
\end{equation}
Here
$x=(x_1,\ldots,x_l)$ denotes independent variables,
$u$ is a dependent variable,
$u_{(n)}$ is the set of all the partial derivatives of the function $u$ with respect to $x$
of order no greater than $n,$ including $u$ as the derivative of zero order.
$L$ is a fixed function of $x,$ $u_{(n)}$ and $\theta.$
$\theta$ denotes the set of of arbitrary (parametric) functions
$\theta(x,u_{(n)})=(\theta^1(x,u_{(n)}),\ldots,\theta^k(x,u_{(n)}))$
satisfying the conditions
\begin{equation}\label{examco}
S(x,u_{(n)},\theta_{(q)}(x,u_{(n)}))=0, \quad S=(S_1,\ldots,S_r).
\end{equation}
These conditions consist of $r$ differential equations on $\theta$,
where $x$ and $u_{(n)}$ play the role of independent variables
and $\theta_{(q)}$ stands for the set of all the partial derivatives of $\theta$ of order no greater than $q$.
In what follows we call the functions $\theta(x,u_{(n)})$ arbitrary elements.
Denote the class of equations of form~(\ref{exameq}) with the arbitrary elements $\theta$
satisfying the constraint~(\ref{examco}) as~$L|_S.$

Let the functions $\theta$ be fixed.
Each one-parametric group of local point transformations
that leaves equation~(\ref{exameq}) invariant corresponds to an
infinitesimal symmetry operator of the form
\[
Q=\xi^a(x,u)\p_{x_a}+\eta(x,u)\p_u.
\]
(here the summation over the repeated indices is understood).
The complete set of such groups generates the principal group
$G^{\max}=G^{\max}(L,\theta)$ of equation~(\ref{exameq}).
The principal group $G^{\max}$ has a corresponding Lie algebra
$A^{\max}=A^{\max}(L,\theta)$ of infinitesimal  symmetry operators
of equation~(\ref{exameq}).
The kernel of principal groups is the group
\[
G^{\ker}=G^{\ker}(L,S)=\bigcap_{\theta:S(\theta)=0}G^{\max}(L,\theta)
\]
for which the Lie algebra is
\[
A^{\ker}=A^{\ker}(L,S)=\bigcap_{\theta:S(\theta)=0}A^{\max}(L,\theta).
\]
Let $G^{\Equiv}=G^{\Equiv}(L,S)$ denote the local transformations group preserving the
form of equations from~$L|_S.$
(Sometimes one consider a subgroup instead the complete equivalence group.)

The problem of group classification consists in finding of all possible
inequivalent cases of extensions of $A^{\max}$, i.e.
in a listing all $G^{\Equiv}$-inequivalent  values of $\theta$ that satisfy
equation~(\ref{examco}) and the condition
$A^{\max}(L,\theta)\ne A^{\ker}$.

In the approach used here group classification is application
of the following algorithm~\mbox{\cite{Ovsiannikov1,Akhatov&Gazizov&Ibragimov1989init}:}
\begin{enumerate}
\item
From the infinitesimal Lie invariance criterion we find the system of
determining equations for the coefficients of $Q$. It is possible that
some of the determining equations does not contain arbitrary elements
and therefore can be integrated immediately. Others
(i.e. the equations containing arbitrary elements explicitly)
are called classifying equations. The main difficulty of group classification
is the need to solve classifying equations with
respect to the coefficients of the operator $Q$ and arbitrary
elements simultaneously.
\item
The next step involves finding of the kernel algebra $A^{\ker}$ of
principal groups of equations from~$L|_S.$
After decomposing the determining equations with respect
to all the unconstrained derivatives of arbitrary elements one obtains a system
of partial differential equations for coefficients of the
infinitesimal operator $Q$ only. Solving of this system yields the algebra $A^{\ker}$.
\item
In order to construct the equivalence group $G^{\Equiv}$ of the class~$L|_S$
we have to investigate the local symmetry transformations
of system~{(\ref{exameq}), (\ref{examco})}, considering it as a system
of partial differential equations with respect to $\theta$ with the
independent variables $x,\ u_{(n)}$. Usually one considers only transformations
being projectible on the space of the variables~$x$ and $u.$
Although in the case when $\theta$ depends, at most, on these variables, it can be assumed
the transformations of them depend on $\theta$ too.
After restricting ourselves to studying of the connected component of unity in $G^{\Equiv},$
we can use the Lie infinitesimal method. To find the complete  equivalence group
(including discrete transformations) we are supposed to use a more complicated direct method.
\item
If $A^{\max}$ is an extension of $A^{\ker}$
(i.e. when $A^{\max}(L,\theta)\ne A^{\ker}$), then
the classifying equations define a system of nontrivial equations for~$\theta$.
Depending on their form and number, we obtain different cases of extensions of~$A^{\ker}$.
To completely integrate the determining equations
we have to investigate a large number of such cases. In order to
avoid cumbersome enumeration of possibilities in solving the determining
equations we can use, for instance, algebraic methods~\cite{Gagnon89c,Zhdanov&Lahno1999,
Basarab-Horwath&Lahno&Zhdanov2001,Lahno&Samoilenko2002-I,Lahno&Samoilenko2002-II,Lahno&Spichak&Stognii2002},
a method which involves the investigation of compatibility of the classifying
equations~\cite{Nikitin,Popovych,Boyko,Vasilenko}
or combined methods~\cite{Popovych&Ivanova&Eshraghi2003CubicLanl,Popovych&Ivanova&Eshraghi2003GammaLanl}.
\end{enumerate}

The result of application of the above algorithm  is a list of equations with their
Lie invariance algebras.
The problem of group classification is assumed to be completely solved if
\begin{enumerate}[\it i\rm)]
\item
the list contains all the possible inequivalent cases of extensions;
\item\vspace{-1ex}
all the equations from the list are mutually inequivalent with respect to
the transformations from $G^{\Equiv}$;
\item\vspace{-1ex}
the obtained algebras are the maximal invariance algebras of the respective equations.
\end{enumerate}
Such list may include equations being mutually equivalent with respect to
local transformations which do not belong to $G^{\Equiv}.$
Knowing such additional equivalences allows to simplify essentially
further investigation of~$L|_S$.
Constructing them can be considered as the fifth step of the algorithm of group classification.
Then, the above enumeration of requirements to the resulting list of classification can be completed
by the following step:
\begin{enumerate}[\it i\rm)]\setcounter{enumi}{3}
\item
all the possible additional equivalences between the listed equations are constructed in
explicit form.
\end{enumerate}
One of the ways for finding additional equivalences is based on the fact that equivalent equations
have equivalent maximal invariance algebras.
The second way is systematical study of conditional and partial
equivalence transformations in the class~$L|_S.$
Let us give definition of such transformations.
Consider a system
\begin{equation}\label{examco1}
S'(x,u_{(n)},\theta_{(q')}(x,u_{(n)}))=0, \quad S'=(S'_1,\ldots,S'_{r'}),
\end{equation}
formed by $r'$ differential equations on $\theta$ with $x$ and $u_{(n)}$ as independent variables.
Let $G^{\Equiv}(L,(S,S'))$ denote the equivalence group of the subclass~$L|_{S,S'}$ of $L|_S,$ where
the functions $\theta$ satisfy systems~(\ref{examco}) and~(\ref{examco1}) simultaneously.

\begin{notion}
We call the transformations from $G^{\Equiv}(L,(S,S'))$
a conditional equivalence transformations of class~$L|_S$ (under the additional constraint $S'$).
The local transformations which transform equations from $L|_{S,S'}$ to $L|_S$
are called partial equivalence transformations of the class~$L|_S$ (under the additional constraint $S'$).
\end{notion}

It is obvious that any conditional equivalence is a partial one under the same additional constraint
and any local symmetry transformation of equation~(\ref{exameq}) for a fixed value $\theta=\theta^0(x,u_{(n)})$
is a partial equivalence transformation under the constraint $\theta=\theta^0.$
The problem of description of all the possible partial equivalence transformations
in the class~$L|_S$ is equivalent to that of local transformations between two
arbitrary equations from $L|_S.$
Additional constraints on arbitrary elements never imply constraining sets of partial equivalences.

\section{Results of classification}\label{sec_roc}
Consider a one-parameter Lie group of local transformations in $(t,x,u)$
with an infinitesimal operator of the form
$
Q=\xi^t(t,x,u)\p_t+\xi^x(t,x,u)\p_x+\eta(t,x,u)\p_u,
$
which keeps equation~(\ref{maineq}) invariant.
The Lie criterion of infinitesimal invariance yields the following determining equations for
$\xi^t,\ \xi^x$ and $\eta$
\begin{equation}\label{dse}
\begin{split}
&\xi^t_x=\xi^t_u=\xi^x_u=0,\\
& D\eta_{uu}+D_u\eta_u-D_u(2\xi^x_x-\xi^t_t)
+D_{uu}\eta-\frac{f_x}f D_u\xi^x=0, \\
&2\xi^x_x-\xi^t_t+\frac{f_x}f\xi^x=\frac{D_u}D \eta ,\qquad
f\eta_t-K\eta_x-D\eta_{xx}=0, \\
&K\left(\frac{f_x}f\xi^x+\xi^x_x-\xi^t_t\right)
+D(\xi^x_{xx}-2\eta_{xu})-2D_u\eta_x-K_u\eta-f\xi^x_t=0.
\end{split}
\end{equation}
Investigating the compatibility of system~(\ref{dse})
we obtain an additional equation $\eta_{uu}=0$
without arbitrary elements.
With this condition system~(\ref{dse})
can be rewritten in the form
\begin{gather}
\xi^t_x=\xi^t_u=\xi^x_u=0, \quad \eta_{uu}=0, \label{dse_modified_0} \\[1ex]
2\xi^x_x-\xi^t_t+\dfrac{f_x}f\xi^x=\dfrac{D_u}D \eta ,
                                               \label{dse_modified_1}  \\[1ex]
D\eta_{xx}+K\eta_x-f\eta_t=0,
                                               \label{dse_modified_2}  \\[1ex]
(D_uK-K_uD)\dfrac{\eta}D - K\xi^x_x-2D_u\eta_x+
              D\xi^x_{xx}-f\xi^x_t-2D\eta_{xu}=0.
                                               \label{dse_modified_3}
\end{gather}

Equations~(\ref{dse_modified_0}) do not contain arbitrary elements.
Integration of them yields
\begin{equation}\label{dse_modified}
\xi^t=\xi^t(t),\quad \xi^x=\xi^x(t,x),\quad \eta=\eta^1(t,x)u+\eta^0(t,x).
\end{equation}

Thus, group classification of~(\ref{maineq}) reduces to solving
of classifying conditions~(\ref{dse_modified_1})--(\ref{dse_modified_3}).

Splitting of system~(\ref{dse_modified_1})--(\ref{dse_modified_3})
with respect to the arbitrary elements and their non-vanishing derivatives gives
the equations $\xi^t_t=0,$ $\xi^x=0,$ $\eta=0$
on the coefficients of operators from $A^{\ker}$ of~(\ref{maineq}).
As a result, the following theorem is true.

\begin{theorem}\label{th1}
The Lie algebra of the kernel of principal groups of~(\ref{maineq}) is
$A^{\ker}=\langle \p_t \rangle.$
\end{theorem}

The next step of algorithm of group classification is finding of
equivalence transformations of class~(\ref{maineq}).
To find these transformations, we have
to investigate Lie symmetries of the system that consists of
equation~(\ref{maineq}) and additional conditions
\[
f_t=f_u=0,\quad D_t=D_x=0,\quad K_t=K_x=0.
\]

Using the classical Lie approach, we find the
invariance algebra of the above system that forms the Lie algebra of $G^{\Equiv}$
for class~(\ref{maineq}). Thus, we obtain the following statement.

\begin{theorem}
The Lie algebra of $G^{\Equiv}$ for class~(\ref{maineq}) is
\begin{equation}\label{eqal}
A^{\Equiv}=\langle \p_t ,\; \p_x ,\; \p_u,\; t\p_t+f\p_f,\;
x\p_x-2f\p_f-K\p_K,\; u\p_u,\; f\p_f+K\p_K+D\p_D\rangle.
\end{equation}
\end{theorem}

Therefore, $G^{\Equiv}$ contains the following continuous transformations:
\[
\begin{split}
&\tilde t=te^{\varepsilon_4}+\varepsilon_1, \
 \tilde x=xe^{\varepsilon_5}+\varepsilon_2, \
 \tilde u=ue^{\varepsilon_6}+\varepsilon_3, \\[1ex]
&\tilde f=fe^{\varepsilon_4-2\varepsilon_5+\varepsilon_7}, \
 \tilde D=De^{\varepsilon_7}, \
 \tilde K=Ke^{-\varepsilon_5+\varepsilon_7},
\end{split}
\]
where $\varepsilon_1,\ldots,\varepsilon_7$ are arbitrary constants.
For class~(\ref{maineq}) there also exists a nontrivial group of
discrete equivalence transformations generated by four involutive
transformations of alternate sign in the sets
$\{t,D,K\},$ $\{x,K\},$ $\{u\}$ and $\{f,D,K\}.$
It can be proved by the direct method that $G^{\Equiv}$ coincides with
the group generated by the both continuous and discrete transformations from the above list.

\begin{theorem}
A complete set of inequivalent equations~(\ref{maineq}) with respect to the transformations from~$G^{\Equiv}$
with $A^{\max}\not=A^{\ker}$
is exhausted by cases given in Tables~\ref{dall}-\ref{dum}.
\end{theorem}

\pagebreak

\newcounter{tbn} \setcounter{tbn}{0}\setcounter{table}{0}
\begin{center}\renewcommand{\arraystretch}{1.1}\refstepcounter{table}\label{dall}
Table~\thetable: Case  $\forall D(u)$\\[2ex]\footnotesize
\begin{tabular}{|l|c|c|l|}
\hline \vspacebefore
N & $K(u)$ & $f(x)$ & \hfill Basis of $A^{\max}\hfill$\\
\hline
\refstepcounter{tbn}\label{t11}
  \thetbn &$\forall$ &$\forall$ &$ \p_t $\\
\refstepcounter{tbn}\label{t12}
  \thetbn a&$\forall$ &$e^{\varepsilon x}$ &$ \p_t,\; \varepsilon t\p_t+\p_x $\\
  \thetbn b&$D$ &$e^{-2x+\gamma e^{-x}}$ &$\p_t,\; \gamma t\p_t-e^x\p_x$\\
  \thetbn c&$D$ &$e^{-2 x}(e^{-x}+\gamma)^\nu$ & $\p_t,\; (\nu+2) t\p_t-(e^{-x}+\gamma)e^x\p_x$\\
  \thetbn d& 0 &$|x|^{\nu}$ &$ \p_t,\; (\nu+2)t\p_t+x\p_x $\\
  \thetbn e&$1 $ &$x^{-1}$ &$ \p_t,\; e^{-t}(\p_t-x\p_x) $\\
\refstepcounter{tbn}\label{t15}
  \thetbn a& 0 & 1 & $ \p_t,\;  \p_x,\;  2t\p_t+x\p_x $\\
  \thetbn b&$D$ &$e^{-2x}$ &$\p_t,\; 2t\p_t-\p_x,\; e^x\p_x$\\
\hline
\end{tabular}
\end{center}
{\footnotesize
Here $\gamma,\nu \ne0,\;$ $\varepsilon=0,1\!\!\!\mod G^{\Equiv},\;$ $\gamma=\pm 1\mod G^{\Equiv}.\;$\\[1ex]
Additional equivalence transformations:\\[1ex]
1. \ref{t12}b $\to$ \ref{t12}a ($K=0,$ $\varepsilon=1$):
 $\tilde t=t,$ $\tilde x=\gamma e^{-x},$ $\tilde u=u;$ \\[1ex]
2. \ref{t12}c $(\nu\ne-2)$ $\to$ \ref{t12}a ($K=-D/(\nu+2),$ $\varepsilon=1$):
    $\tilde t=t,$ $\tilde x=(\nu+2)\ln|e^{-x}+\gamma|,$ $\tilde u=u;$\\[1ex]
$\phantom{\mbox{2. {}}}$\ref{t12}c $(\nu=-2)$ $\to$ \ref{t12}a ($K=-D,$ $\varepsilon=0$):
$\tilde t=t,$ $\tilde x=\ln|e^{-x}+\gamma|,$ $\tilde u=u;$\\[1ex]
3. \ref{t12}d $(\nu\ne-2)$ $\to$ \ref{t12}a ($K=-D/(\nu+2),$ $\varepsilon=1$):
$\tilde t=t,$ $\tilde x=(\nu+2)\ln |x|,$ $\tilde u=u;$\\[1ex]
$\phantom{\mbox{3. {}}}$\ref{t12}d $(\nu=-2)$ $\to$ \ref{t12}a ($K=-D,$ $\varepsilon=0$):
$\tilde t=t,$ $\tilde x=\ln |x|,$ $\tilde u=u;$\\[1ex]
4. \ref{t12}e $\to$ \ref{t12}a ($K=-D,$ $\varepsilon=1$): $\tilde t=e^{t},$ $\tilde x=\ln |x|+t,$ $\tilde u=u;$ \\[1ex]
5. \ref{t15}b $\to$ \ref{t15}a: $\tilde t=t,$ $\tilde x=e^{-x},$ $\tilde u=u.$
\par}

\vspace{2ex}

\setcounter{tbn}{0}
\begin{center}\renewcommand{\arraystretch}{1.1}\refstepcounter{table}\label{dexp}
Table~\thetable: Case $D(u)=e^{\mu u}$\\[2ex]\footnotesize
\begin{tabular}{|l|l|c|c|l|}
\hline \vspacebefore
N & $\mu$ &$K(u)$ & $f(x)$ & \hfill Basis of $A^{\max}\hfill$\\
\hline
\refstepcounter{tbn}\label{t21}
\thetbn&$\forall$&$e^{\nu u}$ & $|x|^{\lambda}$&$\p_t,\;
(\lambda\mu-\lambda\nu+\mu-2\nu)t\p_t+(\mu-\nu)x\p_x+\p_u$\\
\refstepcounter{tbn}\label{t22}
\thetbn&$\forall$&$e^u$ & 1&  $ \p_t,\; \p_x,\; (\mu-2)t\p_t+(\mu-1)x\p_x+\p_u $ \\
\refstepcounter{tbn}\label{t26}
\thetbn&1&$u$ &1 &$ \p_t,\; \p_x,\; t\p_t+(x-t)\p_x+\p_u $\\
\refstepcounter{tbn}\label{t23}\thetbn&1&$\varepsilon e^u$ &$\forall$ &$ \p_t,\; t\p_t-\p_u $\\
\refstepcounter{tbn}\label{t24}
\thetbn a&1&0 &$f^1(x)$ &$ \p_t,\; t\p_t-\p_u,
       \; \alpha t\p_t+(\beta x^2+\gamma_1x+\gamma_0)\p_x+\beta x\p_u$\\
\thetbn b&1&$e^u$ &$f^2(x)$ &$ \p_t,\; t\p_t-\p_u,
      \; \alpha t\p_t-(\beta e^{-x}+\gamma_1+\gamma_0e^x)\p_x+\beta e^{-x}\p_u $\\
\thetbn c&1&$1$ &$x^{-1}$ &$ \p_t,\; x\p_x+\p_u,\; e^{-t}(\p_t-x\p_x)$\\
\refstepcounter{tbn}\label{t25}
\thetbn a&1&0 &1 &$ \p_t,\; t\p_t-\p_u,\; 2t\p_t+x\p_x,\; \p_x$\\
\thetbn b&1&$e^u$ &$e^{-2x}$ & $ \p_t,\; t\p_t-\p_u,\; 2t\p_t-\p_x,\; e^x\p_x $\\
\thetbn c&1&$e^u$ &${e^{-2x}}{(e^{-x}+\gamma)^{-3}}$
      &$ \p_t,\; t\p_t-\p_u,\; (e^{-x}+\gamma)e^x\p_x+\p_u,\;
           -(e^{-x}+\gamma)^2e^x\p_x+(e^{-x}+\gamma)\p_u $\\
\thetbn d&1&0 &$x^{-3}$ &$ \p_t,\; t\p_t-\p_u,\; x\p_x-\p_u,\; x^2\p_x+x\p_u $\\
\hline
\end{tabular}
\end{center}
{\footnotesize
Here $\lambda\ne0,\;$ $\varepsilon\in\{0,1\}\!\!\mod G^{\Equiv},\;$
$\alpha,\beta, \gamma_1, \gamma_0=\const\;$ and
\[
f^1(x)=\exp\left\{\int\frac{-3\beta x-2\gamma_1+\alpha}
        {\beta x^2+\gamma_1x+\gamma_0}\,dx\right \},\qquad
f^2(x)=\exp\left\{\int \dfrac{\beta e^{-x}-2\gamma_0e^x-\alpha}
        {\beta e^{-x}+\gamma_1+\gamma_0e^x}\,dx\right\}.
\]
Additional equivalence transformations:\\[1ex]
1. \ref{t24}b $\to$ \ref{t24}a: $\tilde t=t,$ $\tilde x=e^{-x},$ $\tilde u=u;$ \\[1ex]
2. \ref{t24}c $\to$ \ref{t24}a ($\alpha=\gamma_0=1,$ $\beta=\gamma_1=0,$ $f^1=x^{-1}$):
  $\tilde t=e^t,$ $\tilde x=e^tx,$ $\tilde u=u;$ \\[1ex]
3. \ref{t25}b $\to$ \ref{t25}a: $\tilde t=t,$ $\tilde x=e^{-x},$ $\tilde u=u;$ \\[1ex]
4. \ref{t25}c $\to$ \ref{t25}a: $\tilde t=t\sign (e^{-x}+\gamma),$ $\tilde x=1/(e^{-x}+\gamma),$
                             $\tilde u=u-\ln|e^{-x}+\gamma|;$\\[1ex]
5. \ref{t25}d $\to$ \ref{t25}a: $\tilde t= t\sign x,$ $\tilde x=1/x,$ $\tilde u=u-\ln |x|.$
\par}

\pagebreak

\setcounter{tbn}{0}
\begin{center}\renewcommand{\arraystretch}{1.1}\refstepcounter{table}\label{dum}
Table~\thetable: Case $D(u)=u^{\mu}$\\[2ex]\footnotesize
\begin{tabular}{|l|c|c|c|l|}
\hline \vspacebefore
N & $\mu$ & $K(u)$ & $f(x)$ & \hfill Basis of $A^{\max}\hfill$\\
\hline
\refstepcounter{tbn}\label{t31}
\thetbn&$\forall$ &$u^\nu$
&$|x|^\lambda$ &$ \p_t,\; (\mu+\lambda\mu-2\nu-\lambda\nu)t\p_t+(\mu-\nu)x\p_x+u\p_u $\\
\refstepcounter{tbn}\label{t32}
\thetbn&$\forall$ &$u^\nu$&1 &$ \p_t,\; \p_x,\; (\mu-2\nu)t\p_t+(\mu-\nu)x\p_x+u\p_u $\\
\refstepcounter{tbn}\label{t312}
\thetbn &$\forall$ &$\ln u $ &$1$ &$\p_t,\;\p_x,\; \mu t\p_t+(\mu x-t)\p_x+u\p_u  $ \\
\refstepcounter{tbn}\label{t33}
\thetbn&$\forall$ &$\varepsilon u^\mu$ &$\forall$ &$ \p_t,\; \mu t\p_t-u\p_u $\\
\refstepcounter{tbn}\label{t34}
\thetbn a&$\forall$ &0 &$f^3(x)$ &$ \p_t,\; \mu t\p_t-u\p_u,$ \\
&&&& $\alpha t\p_t+((1+\mu)\beta x^2+\gamma_1x+\gamma_0)\p_x+\beta xu\p_u$\\
\thetbn b&$\forall$ &$u^\mu$ &$f^4(x)$&$ \p_t,\; \mu t\p_t-u\p_u,$\\
&&&&$\alpha t\p_t-((1+\mu)\beta e^{-x}+\gamma_1+\gamma_0e^x)\p_x+\beta e^{-x}u\p_u$\\
\thetbn c&$\mu\ne-3/2$ &$1$ &$x^{-1}$&$ \p_t,\; e^{-t}(\p_t-x\p_x),\; \mu x\p_x+u\p_u$\\
\refstepcounter{tbn}\label{t36}
\thetbn a&$\mu \ne -4/3$ &0 &1 &$ \p_t,\; \mu t\p_t-u\p_u,\; \p_x,\; 2t\p_t+x\p_x$\\
\thetbn b&$\mu\ne -4/3$ &$u^{\mu}$ &$e^{-2x}$&$\p_t,\; \mu t\p_t-u\p_u,\; 2t\p_t-\p_x,\; e^x\p_x$\\
\thetbn c&$-1$ &0 &$e^{\gamma x}$
   &$ \p_t,\; t\p_t+u\p_u,\; \p_x-\gamma u\p_u,\;2t\p_t+x\p_x-\gamma xu\p_u $\\
\thetbn d&$-1$ &$u^{-1}$ &$e^{-2x+\gamma e^{-x}}$
    &$ \p_t,\; t\p_t+u\p_u,\; e^x\p_x+\gamma u\p_u,\;2t\p_t-\p_x-\gamma e^{-x}u\p_u $\\
\thetbn e&$\mu \ne -4/3,-1$ &0 &$|x|^{-\frac{4+3\mu}{1+\mu}}$
   &$\p_t,\; \mu t\p_t-u\p_u,\;  (2+\mu)t\p_t-(1+\mu)x\p_x,$\\
   &&&& $(1+\mu)x^2\p_x+xu\p_u $\\
\thetbn f&$\mu\ne -4/3,-1$ &$u^{\mu}$
   &$\frac{e^{-2x}}{(e^{-x}+\gamma)^{\frac{4+3\mu}{1+\mu}}}$
   &$ \p_t,\; \mu t\p_t-u\p_u,\; (2+\mu)t\p_t+(1+\mu)(e^{-x}+\gamma)e^x\p_x,$\\
   &&&&$-(1+\mu)(e^{-x}+\gamma )^2e^x\p_x+(e^{-x}+\gamma)u\p_u$\\
\thetbn g&$-3/2$ &$1$ &$x^{-1}$&$ \p_t,\; e^{-t}(\p_t-x\p_x),\; 3 x\p_x-2u\p_u,\; e^t(x^2\p_x-2xu\p_u) $\\
\refstepcounter{tbn}\label{t37}\thetbn a&$-4/3$ &0 &1
   &$ \p_t,\; 4t\p_t+3u\p_u,\; \p_x,\; 2t\p_t+x\p_x,\; x^2\p_x-3xu\p_u $\\
\thetbn b&$-4/3$ &$u^{-4/3}$ &$e^{-2x}$
   &$ \p_t,\; 4t\p_t+3u\p_u,\; 2t\p_t-\p_x,\; e^{-x}(\p_x+3u\p_u),\; e^x\p_x $\\
\refstepcounter{tbn}\label{t313}\thetbn &0&$u$ & 1
   & $ \p_t,\; \p_x,\; 2t\p_t+x\p_x-u\p_u,\; t\p_x-\p_u,$ \\
   &&&&$t^2\p_t+tx\p_x-(tu+x)\p_u $\\
\hline
\end{tabular}
\end{center}

{\footnotesize
Here $\mu\ne0$ for Cases 4--6. $\;\varepsilon=0,1\!\!\!\mod G^{\Equiv},\;$ $\lambda\ne0,\:$
$\alpha,\beta,\gamma_1,\gamma_0=\const\;$ and
\[
f^3(x)=\exp\left\{\int\frac{-(4+3\mu)\beta x-2\gamma_1+\alpha}
        {(1+\mu)\beta x^2+\gamma_1x+\gamma_0}\,dx\right \},\qquad
f^4(x)=\exp\left\{\int \dfrac{(2+\mu)\beta e^{-x}-2\gamma_0e^x-\alpha}
        {(1+\mu)\beta e^{-x}+\gamma_1+\gamma_0e^x}\,dx\right\}.
\]
Additional equivalence transformations:\\[1ex]
1. \ref{t34}b $\to$ \ref{t34}a: $\tilde t=t,$ $\tilde x=e^{-x},$ $\tilde u=u;$ \\[1ex]
2. \ref{t34}c $\to$ \ref{t34}a ($\alpha=\gamma_0=1,$ $\beta=\gamma_1=0,$ $f^1=x^{-1}$):
    $\tilde t=e^t,$ $\tilde x=e^tx,$ $\tilde u=u;$ \\[1ex]
3. \ref{t36}b $\to$ \ref{t36}a: $\tilde t=t,$ $\tilde x=e^{-x},$ $\tilde u=u;$ \\[1ex]
4. \ref{t36}c $\to$ \ref{t36}a ($\mu=-1$): $\tilde t=t,$ $\tilde x=x,$ $\tilde u=e^{\gamma x}u;$\\[1ex]
5. \ref{t36}d $\to$ \ref{t36}a ($\mu=-1$): $\tilde t=t,$ $\tilde x=e^{-x},$ $\tilde u=e^{\gamma e^{-x}}u;$\\[1ex]
6. \ref{t36}e $\to$ \ref{t36}a: $\tilde t=t,$ $\tilde x=-1/x,$ $\tilde u=|x|^{-\frac1{1+\mu}}u;$\\[1ex]
7. \ref{t36}f $\to$ \ref{t36}a: $\tilde t=t,$ $\tilde x=-1/(e^{-x}+\gamma),$
    $\tilde u=|e^{-x}+\gamma|^{-\frac 1{1+\mu}}u;$\\[1ex]
8. \ref{t36}g $\to$ \ref{t36}a ($\mu=-3/2$): $\tilde t=e^t,$ $\tilde x=-e^{-t}/x,$ $\tilde u=|e^tx|^{-\frac1{1+\mu}}u;$ \\[1ex]
9. \ref{t37}b $\to$ \ref{t37}a: $\tilde t=t,$ $\tilde x=e^{-x},$ $\tilde u=u.$ \par
}

\pagebreak

In Tables~\ref{dall}--\ref{dum} we list all possible $G^{\Equiv}$-inequivalent
sets of functions $f(x),$ $D(u),$ $K(u)$ and corresponding invariance algebras.
Numbers with the same arabic figures correspond to cases that are equivalent
with respect to a local equivalence transformation. Explicit formulas for these
transformations are adduced after the Tables.
The cases numbered with different arabic figures are inequivalent
with respect to local equivalence transformations.
In order to simplify the presented results, in the case $f(x)=1$ we just use
the conditional equivalence transformation
$\tilde x=x-\varepsilon t,$ $ \tilde K=K+\varepsilon$
(the other variables are not transformed) from $G^{\Equiv}_1$ (see Section~\ref{sec_f1}).
Other conditional equivalence transformations are considered in Section~\ref{sec.cets}.

Below for convenience we use double numeration T.N of classification cases and local equivalence
transformations, where T denotes the number of table and N does the number of case (or transformation)
in Table~T. The notion ``equation~T.N'' is used for the equation of form~(\ref{maineq}) where
the parameter-functions take values from the corresponding case.

The operators from Tables~\ref{dall}--\ref{dum} form bases of the maximal invariance algebras
iff the corresponding sets of the functions $f,$ $D,$ $K$ are
$G^{\Equiv}$-inequivalent to ones with most extensive invariance algebras.
For example, in Case~3.1 $(\mu,\nu)\ne (0,0)$ and $\lambda \ne -1$ if $\nu=0.$
And in Case~3.2 $(\mu,\nu)\notin \{(-2,-2),(0,1)\}$ and $\nu\ne0.$
Similarly, in Case~\ref{dexp}.1 the constraint set on  the parameters $\mu,$ $\nu$ and $\lambda$
coincides with the one for Case~~\ref{dum}.1, and we can assume that $\mu=1$ or $\nu=1.$
In Case~\ref{dexp}.2 we consider $\nu=1$ immediately.

After analyzing the obtained results, we can state the following theorem.

\begin{theorem}
If an equation of form~(\ref{maineq}) is invariant with respect to a Lie algebra
of dimension no less than 4 then it can be reduced by means of local transformations to one with $f(x)=1.$
\end{theorem}

\section{Group classification for subclass with
\mathversion{bold}$f(x)=1$\mathversion{normal}}\label{sec_f1}

Class~(\ref{maineq}) includes a subclass of equations of the
general form
\begin{equation} \label{maineq1}
u_t=\left(D(u)u_x \right)_x + K(u)u_x.
\end{equation}
(i.e. the function $f$ is assumed to be equal to 1 identically).
Symmetry properties of equations~(\ref{maineq1}) were studied in
\cite{Oron&Rosenau,Edwards}. But we are not aware of any paper containing
correct and exhaustive investigation on the subject. Now let us single out the
results of group classification of equations~(\ref{maineq1}) from the above
section.

\begin{theorem}
The Lie algebra of the kernel of principal groups of~(\ref{maineq1}) is
$A^{\ker}_1=\langle \p_t , \p_x \rangle.$
\end{theorem}

\begin{theorem}
The Lie algebra of the equivalence group $G^{\Equiv}_1$ for the class~(\ref{maineq1}) is
\begin{equation}\label{eqal1}
A^{\Equiv}_1=\langle \p_t ,\; \p_x,\; \p_u,\; u\p_u,\; t\p_x-\p_K,\;
2t\p_t+x\p_x-K\p_K,\; t\p_t-D\p_D-K\p_K\rangle .
\end{equation}
\end{theorem}
Any transformation from $G^{\Equiv}_1$ has the form:
\begin{equation}\label{eqgr1}
\begin{split}
&\tilde t=t\varepsilon_4^2\varepsilon_5+\varepsilon_1, \quad
 \tilde x=x\varepsilon_4+\varepsilon_7 t+\varepsilon_2, \quad
 \tilde u=u\varepsilon_6+\varepsilon_3, \\
&\tilde D=D\varepsilon_5^{-1}, \quad
 \tilde K=K\varepsilon_4^{-1}\varepsilon_5^{-1}-\varepsilon_7,
\end{split}
\end{equation}
where $\varepsilon_1,\ldots,\varepsilon_7$ are arbitrary constants,
$\varepsilon_4\varepsilon_5\varepsilon_6\ne0.$
\begin{theorem}
The complete set of $G^{\Equiv}_1$-inequivalent extensions of $A^{\max}$
for equations~(\ref{maineq1}) is given in Table~\ref{f1}.
\end{theorem}

\setcounter{tbn}{0}
\begin{table} \footnotesize
\caption{Case $f(x)=1.$}
\vspace{-2ex}
\begin{center}\renewcommand{\arraystretch}{1.1}
\begin{tabular}{|l|c|c|l|}
\hline\vspacebefore
N &$D(u)$ & $K(u)$ & \hfill Basis of $A^{\max}\hfill$ \\
\hline\vspacebefore
\refstepcounter{tbn}\label{t51}\thetbn&$\forall$ & $\forall$ & $\p_t,\;\p_x  $
\\
\refstepcounter{tbn}\label{t52}\thetbn&$\forall$ &0 &$\p_t,\;\p_x,\;2t\p_t+x\p_x$
\\
\refstepcounter{tbn}\label{t53}\thetbn&$e^{\mu u}$ &$e^u$ &$\p_t,\;\p_x,\;
                             (\mu-2)t\p_t+(\mu-1)x\p_x+\p_u$
\\
\refstepcounter{tbn}\label{t54}\thetbn&$e^u$ & $u$ &
   $ \p_t,\;\p_x,\;t\p_t+(x-t)\p_x+\p_u  $
\\
\refstepcounter{tbn}\label{t55}\thetbn&$e^u$ & 0 &
   $ \p_t,\;\p_x,\;t\p_t-\p_u,2t\p_t+x\p_x  $
\\
\refstepcounter{tbn}\label{t56}\thetbn&$u^{\mu}$ & $u^\nu$ &
   $ \p_t,\;\p_x,\;(\mu -2\nu)t\p_t+(\mu -\nu)x\p_x+u\p_u  $
\\
\refstepcounter{tbn}\label{t57}\thetbn a&$u^{\mu}$ & 0 &
   $ \p_t,\;\p_x,\;\mu t\p_t-u\p_u,2t\p_t+x\p_x  $
\\
   \thetbn b&$u^{-2}$ & $u^{-2}$ &$ \p_t,\;\p_x,\;2t\p_t+u\p_u,e^{-x}(\p_x+u\p_u)  $
\\
 \refstepcounter{tbn}\label{t58}\thetbn&$u^{-4/3}$ & 0 &
   $ \p_t,\;\p_x,\;4t\p_t+3u\p_u,2t\p_t+x\p_x,\;x^2\p_x-3xu\p_u  $
\\
\refstepcounter{tbn}\label{t59}\thetbn&$u^\mu$ & $\ln u$ &
   $ \p_t,\;\p_x,\;\mu t\p_t+(\mu x-t)\p_x+u\p_u  $
\\
\refstepcounter{tbn}\label{t510}\thetbn&1 & $u$ & $ \p_t,\;\p_x,\;
              t^2\p_t+tx\p_x-(tu+x)\p_u,\;2t\p_t+x\p_x-u\p_u,\; t\p_x-\p_u  $
\\
\hline
\end{tabular}
\end{center}
Here $\mu,\nu=\const$. $(\mu,\nu)\not=(-2,-2),\;(0,1)$ and
$\nu\not=0$ for ${\rm N}=6.$ $\mu\not=-4/3$ for ${\rm N}={\rm 7a}.$
Case~\ref{t57}b can be reduced to~\ref{t57}a ($\mu=-2$) by means of the conditional
equivalence transformation $\tilde t=t,$ $\tilde x=e^x,$ $\tilde u=e^{-x}u.$
\label{f1}
\end{table}

\section{Proof of Theorem~3}\label{sec_proof}

Our method is based on the fact that the substitution of the coefficients of any operator from
$A^{\max} \backslash A^{\ker}$ into the classifying equations results in nonidentity equations for
arbitrary elements.
In the problem under consideration, the procedure of looking for the possible cases mostly
depends on equation~(\ref{dse_modified_1}).
For any operator $Q\in A^{\max}$ equation (\ref{dse_modified_1})
gives some equations on~$D$ of the general form
\begin{equation} \label{deq}
(au+b)D_u=cD,
\end{equation}
where $a,b,c=\const$. In general for all operators from $A^{\max}$
the number $k$ of such independent equations is no greater then 2 otherwise
they form an incompatible system on $D$. $k$ is an invariant value for the
transformations from $G^{\Equiv}$. Therefore, there exist three
inequivalent cases for the value of $k$: $k=0,$ $k=1,$ $k=2.$
Let us consider these possibilities in more detail,
omitting cumbersome calculations.

\medskip

\noindent{\mathversion{bold}$k=0$} (Table~\ref{dall}).
Then the coefficients of any operator from $A^{\max}$ are to satisfy the system
\begin{equation}\label{dse_dall}
\eta=0,\quad 2\xi^x_x-\xi^t_t+\frac{f_x}f\xi^x=0,\quad -K\xi^x_x+D\xi^x_{xx}-f\xi^x_t=0.
\end{equation}

Let us suppose that $K\notin \langle 1,\; D \rangle.$
It follows from the last equation of the system~(\ref{dse_dall}) that $\xi^x_x=\xi^x_t=0.$
Therefore, the second equation is a nonidentity
equation for $f$ of the form $f_x=\mu f$ without fail.
Solving this equation yields Case~2a.

Now let $K\in \langle 1,\; D \rangle,$ i.e. $K=\varepsilon D+\beta$ where
$\varepsilon\in\{0,1\},$ $\beta=\const.$
Then the last equation of~(\ref{dse_dall}) can be decomposed into the following ones
\[
\xi^x_{xx}=\varepsilon \xi^x_x,\quad  \beta\xi^x_x+f\xi^x_t=0.
\]
The equation $(\xi^x(f_x/f+2\varepsilon))_x=0$ is a differential consequence
of the reduced determining equations. Therefore, the condition $f_x/f+2\varepsilon=0$ is
a classifying one.

Suppose this condition is true, i.e. $f=e^{-2\varepsilon x}\mod G^{\Equiv}.$
There exist three different possibilities for values of the parameters $\varepsilon$ and  $\beta$:
\[
\varepsilon=1,\ \beta\ne0;\quad \varepsilon=1,\ \beta=0;\quad
\varepsilon=0\ (\mbox{then}\  \beta=0\!\!\!\mod G^{\Equiv}_1),
\]
which yield Cases~2a, 3b and 3a respectively.

Let $\varepsilon=0$ and $f_x/f\ne 0.$ Then either our consideration is reduced to Case~2a
or $f=x^\mu\!\!\!\mod G^{\Equiv}$ where $\mu\ne0.$ Depending on the value of the parameter $\beta$
($\beta=0$ or $\beta\ne0$ and then $\mu=-1$) we obtain Case~2d or Case~2e.

Let $\varepsilon=1$ and $f_x/f\ne -2.$ Then $\beta=0$ and $f_x/f=(C_1e^x+C_0)^{-1}-2$
where we assume $C_1\ne0$ to exclude Case~2a. Integrating the latter equation depends on
whether $C_0$ vanishes or not,
and results in Cases~2b and~2c.

\medskip

\noindent{\mathversion{bold}$k=1.$ }
Then $D\in\{e^u,u^{\mu},\mu\ne0\}\!\!\mod G^{\Equiv}$
and there exists $Q\in A^{\max}$ with $\eta\ne 0.$

\medskip

Let us investigate the first possibility $D=e^u$  (Table~2).
Equation~(\ref{dse_modified_1}) implies $\eta_u=0,$ i.e. $\eta=\eta(t,x).$
Therefore, equations~(\ref{dse_modified_1})--(\ref{dse_modified_3}) can be written as
\begin{equation}\label{dse_dexp}
\begin{split}
& 2\xi^x_x-\xi^t_t+\dfrac{f_x}f\xi^x=\eta, \qquad
e^u\eta_{xx}+K\eta_x-f\eta_t=0, \\
& (K-K_u)\eta-K\xi^x_x-f\xi^x_t+e^u(\xi^x_{xx}-2\eta_x)=0.
\end{split}
\end{equation}
The latter equation looks  like $K_u=\nu K+be^u+c$ with respect to $K$,
where $\nu,b,c=\const$. Therefore, $K$ is to take one of the following five values.

\medskip

\noindent
1. $K=e^{\nu u}+\vk_1e^u+\vk_0\mod G^{\Equiv},$ where $\nu\ne0,1.$
(Here and below $\vk_i=\const,$ $i=0,1.$)
Then  $\eta=\const,$ $\vk_1=0,$
and either $\vk_0=0$ if $f \ne\const$ or $\vk_0=0\!\!\mod G^{\Equiv}_1$ if $f=\const$
that imply $\xi^x_t=0,$ $\xi^t_{tt}=0,$ therefore $f=|x|^\lambda\!\!\mod G^{\Equiv}$ (Cases~1 and~2).

\medskip

\noindent
2. $K=u+\vk_1e^u+\vk_0.$
In the way analogous to the previous case we obtain
$\vk_1=0,$ $f=1\!\!\mod G^{\Equiv},$ $\vk_0=0\!\!\mod G^{\Equiv}_1$ (Case~3).

\medskip

\noindent
3. $K=ue^u+\vk_1e^u+\vk_0\mod G^{\Equiv}.$
It follows from system~(\ref{dse_dexp}) that $\eta=0$ for any operator from $A^{\max},$
i.e. we have a contradiction to assumption $\eta\ne 0$ for some operator from~$A^{\max}.$

\medskip

\noindent
4. $K=e^u+\vk_0.$
Then $\eta^1=\zeta^1(t)e^{-x}+\zeta^0(t),$ $\xi^x=\sigma^1(t)e^x+\sigma^0(t)-\zeta^1(t)e^{-x}.$
It can be proved that $\zeta^1_t=\zeta^0_t=\sigma^1_t=\xi^t_{tt}=0,$
either $\vk_0=0$ if $f \ne\const$ or $\vk_0=0\!\!\mod G^{\Equiv}_1$ if $f=\const,$
and therefore $\sigma^0_t=0.$
The first equation of~(\ref{dse_dexp}) implies that the function~$f$ is to satisfy
$l$~($l=0,1,2$) equations of the form
\[
\frac{f_x}f=\frac{\beta e^{-x}-\alpha-2\gamma_0 e^x}{\beta e^{-x}+\gamma_1+\gamma_0 e^x}
\]
with non-proportional sets of constant parameters $(\alpha,\beta,\gamma_0,\gamma_1).$
The values $l=0$ and $l=1$ correspond to Cases~4 ($\varepsilon=1$) and~5b.
An additional extension of $A^{\max}$ exists for $l=2$ in comparison with $l=1$
iff $f$ is a solution of the equation
\[
\frac{f_x}f=\frac{\lambda_2 e^{-x}}{\lambda_1 e^{-x}+\lambda_0}-2,
\]
where either $\lambda_2=0$ or $\lambda_2=3\lambda_1\ne 0.$
Integrating the latter equation gives Cases~6b and 6c.

\medskip

\noindent
5. $K=\vk_0.$
Then $\eta^1=\zeta^1(t)x+\zeta^0(t),$ $\xi^x=\sigma^1(t)x+\sigma^0(t)+\zeta^1(t)x^2.$
It follows from compatibility of system~(\ref{dse_dexp}) that
$\eta_t=\xi^x_t=0$ if $f\not\in\{x^{-1},1\}\!\!\mod G^{\Equiv}$ or  $\vk_0=0.$
The values $f=x^{-1},$ $\vk_0\ne 0$ result in Case~5c.
If $f\not\in\{x^{-1},1\}\!\!\mod G^{\Equiv}$ and  $\vk_0=0$, we obtain Case~1 with $\nu=0.$
If $f=\const$ then $\vk_0=0\!\!\mod G^{\Equiv}_1.$
Below $\vk_0=0.$
The first equation of~(\ref{dse_dexp}) holds when the function~$f$ is a solution of a system of
$l$~($l=0,1,2$) equations of the form
\[
\frac{f_x}f=\frac{-3\beta x+\alpha-2\gamma_1}{\beta x^2+\gamma_1 x+\gamma_0}
\]
with non-proportional sets of constant parameters $(\alpha,\beta,\gamma_0,\gamma_1).$
The values $l=0$ and $l=1$ correspond to Cases~4 ($\varepsilon=0$) and~5a.
Additional extensions for $l=2$ exist iff $f$ is a solution of the equation
\[
\frac{f_x}f=\frac{\lambda_2}{\lambda_1 x+\lambda_0}.
\]
where either $\lambda_2=0$ or $\lambda_2=-3\lambda_1\ne 0.$
These possibilities result in Cases~6a and~6d.

\medskip

Consider the case $D=u^{\mu}$ (Tables~3).
Equation~(\ref{dse_modified_1}) implies $\eta^0=0,$ i.e. $\eta=\eta^1(t,x)u.$
Therefore, system~(\ref{dse_modified_1})--(\ref{dse_modified_3}) can be written as
\begin{equation}\label{dse_dum}
\begin{split}
&2\xi^x_x-\xi^t_t+\dfrac{f_x}f\xi^x=\mu\eta^1 ,\qquad
u^\mu\eta^1_{xx}+K\eta^1_x-f\eta^1_t=0,\\
&(\mu K-uK_u)\eta^1 - K\xi^x_x+(\xi^x_{xx}-2(\mu+1)\eta^1_x)u^\mu-f\xi^x_t=0.
\end{split}
\end{equation}
The latter equation looks with respect to $K$ similarly to $uK_u=\nu K+bu^\mu+c,$
where $\nu,b,c=\const$. Therefore, $K$ is to take one of the five values.

\medskip

\noindent
1. $K=u^\nu+\vk_1u^{\mu}+\vk_0\mod G^{\Equiv},$ where $\nu\ne0,\mu.$
Equations~(\ref{dse_dum}) imply $\eta^1=\const,$ $\xi^x=(\mu-\nu)\eta^1x+\sigma(t),$
$\vk_1\xi^x_x=0$ (therefore, $\vk_1=0$ since $\eta^1=0$), $f=|x|^\lambda\!\!\mod G^{\Equiv},$
$\xi^t_t=(\mu+\lambda\mu-2\nu-\lambda\nu)\eta^1,$ $\lambda\sigma=0,$
and either $\vk_0=0$ if $\lambda\ne0$ (Case~1) or $\vk_0=0\!\!\mod G^{\Equiv}_1$ if $\lambda=0$ (Case~2).

\medskip

\noindent
2. $K=\ln u+\vk_1 u^{\mu}+\vk_0\mod G^{\Equiv}.$
In the way analogous to the previous case we obtain
$\vk_1=0,$ $f=1\!\!\mod G^{\Equiv},$ $\vk_0=0\!\!\mod G^{\Equiv}_1$ (Case~3).

\medskip

\noindent
3. $K=u^{\mu}\ln u+\vk_1 u^{\mu}+\vk_0\mod G^{\Equiv}.$
It follows from system~(\ref{dse_dum}) that $\eta=0$ for any operator from $A^{\max},$
i.e. we have a contradiction to assumption $\eta\ne 0$ for some operator from~$A^{\max}.$

\medskip

\noindent
4. $K=u^{\mu}+\vk_0\mod G^{\Equiv}.$
Then $\eta^1=\zeta^1(t)e^{-x}+\zeta^0(t),$ $\xi^x=\sigma^1(t)e^x+\sigma^0(t)-(\mu+1)\zeta^1(t)e^{-x}.$
It can be proved that $\zeta^1_t=\zeta^0_t=\sigma^1_t=\xi^t_{tt}=0,$
either $\vk_0=0$ if $f \ne\const$ or $\vk_0=0\!\!\mod G^{\Equiv}_1$ if $f=\const,$
and therefore $\sigma^0_t=0.$
The first equation of~(\ref{dse_dum}) implies that the function~$f$ is to satisfy
$l$~($l=0,1,2$) equations of the form
\[
\frac{f_x}f=\frac{(\mu+2)\beta e^{-x}-\alpha-2\gamma_0 e^x}{(\mu+1)\beta e^{-x}+\gamma_1+\gamma_0 e^x}
\]
with non-proportional sets of constant parameters $(\alpha,\beta,\gamma_0,\gamma_1).$
The values $l=0$ and $l=1$ correspond to Cases~4 ($\varepsilon=1$) and~5b.
$l=2$ iff $f$ is a solution of the equation
\[
\frac{f_x}f=\frac{\lambda_2 e^{-x}}{\lambda_1 e^{-x}+\lambda_0}-2.
\]
Looking for the inequivalent possibilities of integrating of this equation results in
Cases~6b, 6d, 6f, 7b.

\medskip

\noindent
5. $K=\vk_0.$
Then $\eta^1=\zeta^1(t)x+\zeta^0(t),$ $\xi^x=\sigma^1(t)x+\sigma^0(t)+(\mu+1)\zeta^1(t)x^2.$
It follows from compatibility of system~(\ref{dse_dum}) that
$\eta_t=\xi^x_t=0$ if $f\not\in\{x^{-1},1\}\!\!\mod G^{\Equiv}$ or  $\vk_0=0.$
The values $f=x^{-1},$ $\vk_0\ne 0$ result in Cases~5c and~6g.
If $f\not\in\{x^{-1},1\}\!\!\mod G^{\Equiv}$ and  $\vk_0=0$, we obtain Case~1 with $\nu=0.$
If $f=\const$ then $\vk_0=0\!\!\mod G^{\Equiv}_1.$
Below $\vk_0=0.$
The first equation of~(\ref{dse_dum}) holds when the function~$f$ is a solution of a system of
$l$~($l=0,1,2$) equations of the form
\[
\frac{f_x}f=\frac{-(3\mu+4)\beta x+\alpha-2\gamma_1}{(\mu+1)\beta x^2+\gamma_1 x+\gamma_0}
\]
with non-proportional sets of constant parameters $(\alpha,\beta,\gamma_0,\gamma_1).$
The values $l=0$ and $l=1$ correspond to Cases~4 ($\varepsilon=0$) and~5a.
$l=2$ iff $f$ is a solution of the equation
\[
\frac{f_x}f=\frac{\lambda_2}{\lambda_1 x+\lambda_0}.
\]
Looking for the inequivalent possibilities of integrating of this equation results in
Cases~6a, 6c, 6e, 7a.

\medskip

\noindent\mathversion{bold}\mbox{$k=2.$}\mathversion{normal}
Assumption on two independent equations of form~(\ref{deq}) on $D$ yields
$D=\const,$ i.e. $D=1\mod G^{\Equiv}.$ $K_u\not=0$ (otherwise, equation~(\ref{maineq}) is linear).
Equations~(\ref{dse_modified_1})--(\ref{dse_modified_3}) can be written as
\begin{equation} \label{dse_d1}
\begin{split}
&2\xi^x_x-\xi^t_t+\dfrac{f_x}f\xi^x=0, \qquad
\eta_{xx}+K\eta_x-f\eta_t=0,\\
&-K_u\eta-K\xi^x_x+\xi^x_{xx}-f\xi^x_t-2\eta^1_x=0.
\end{split}
\end{equation}
The latter equation looks similarly to $(au+b)K_u=cK+d$ with respect to $K,$
where $a,b,c,d=\const$. Therefore, to within transformations from $G^{\Equiv}$
$K$ is to take one from four values:
\[
K=u^\nu+\vk_0,\;\nu\ne0,1;\quad
K=\ln u+\vk_0;\quad
K=e^u+\vk_0;\quad
K=u.
\]
Classification for these values is carried out in the way similar to the above.
The obtained extensions can be entered in either Table~2 or Table~3.

The problem of the group classification of equation~(\ref{maineq}) is
exhaustively solved.

\section{Additional equivalence transformations}\label{sec.cets}

When we impose some restrictions on arbitrary elements we can find additional equivalence
transformations named conditional equivalence transformations (see Notion~1).
As mentioned above, the simplest way to find such equivalences between previously
classified equations is based on the fact that equivalent equations have equivalent maximal invariance algebras.
A more systematic way is to classify these transformations using the infinitesimal or the direct methods.
Examples of conditional equivalence algebras calculated by the infinitesimal method
are listed in Table~\ref{tableequiv}.

\begin{table} \footnotesize
\caption{Conditional equivalence algebras}
\begin{center}\renewcommand{\arraystretch}{1.3}
\begin{tabular}{|l|l|}
\hline \vspacebefore
\hfill {Conditions \hfill}& \hfill Basis of $A^{\Equiv}\hfill$\\
\hline
$K=D$ & $\p_t,\;\p_x,\; \p_u,\; u\p_u,\; t\p_t+f\p_f,\; e^x(\p_x-2f\p_f),\; f\p_f+D\p_D$\\
$K=D=e^u $& $ \p_t,\ t\p_t+f\p_f,\  \p_x,\  \p_u+f\p_f,\  x\p_x-2f\p_f,\ x^2\p_x+x\p_u-3x\p_f$\\
$D=e^u,$ $K=0$&$\p_t,\ t\p_t+f\p_f,\ \p_x,\ \p_u+f\p_f,\ x\p_x-2f\p_f,\ x^2\p_x+x\p_u-3x\p_f $\\
$D=K=u^{\mu}$&$ \p_t,\; t\p_t+f\p_f,\; \p_x,\; \p_u+\mu f\p_f,\; e^x(\p_x-2f\p_f),\;
e^{-x}((1\!+\!\mu)\p_x-u\p_u+(2\!+\!\mu)f\p_f)$\\
$D=u^{\mu},$ $K=0$ &$\p_t,\ t\p_t+f\p_f,\ \p_x,\ \p_u+\mu f\p_f,\ x\p_x-2f\p_f,\ (1+\mu)x^2\p_x+xu\p_u-(4+3\mu)xf\p_f$\\
\hline
\end{tabular}
\end{center}
\label{tableequiv}
\end{table}

To find the complete collection of additional local equivalence transformations including
both continuous and discrete ones, we are to use the direct method.
Moreover, application of this method allows us to describe all the local transformations
that are possible for pairs of equations from the class under consideration.
The problem of this sort was first investigated for wave equations by Kingston and
Sophocleous~\cite{Kingston98, Sophocleous99, Kingston01}.
Now we state a number of simple but very useful lemmas containing preliminary results on
solution of this problem.
(We imply that the condition of nonsingularity is satisfied for all the transformations.)

\begin{lemma}\cite{Kingston91}
For any local transformation between two evolutionary second-order equations
(i.e. equations of the form $u_t=H(t,x,u,u_x,u_{xx})$ where $H_{u_{xx}}\ne 0$)
the transformation of the variable $t$ depends only on $t.$
\end{lemma}

\begin{lemma}
Any local transformation between two evolutionary second-order quasi-linear equations having the form
$u_t=F(t,x,u)u_{xx}+G(t,x,u,u_x)$  where $F\ne 0$ is projectible, i.e.
$\tilde t =T(t),$ $\tilde x =X(t,x),$ $\tilde u =U(t,x,u).$
\end{lemma}

\begin{lemma}
Any local transformation between two equations from class~(\ref{maineq}) is linear with respect to $u$:
$\tilde t =T(t),$ $\tilde x =X(t,x),$ $\tilde u =U^1(t,x)u+U^0(t,x),$
and up to transformations from $G^{\Equiv}$ we can assume the coefficient~$D$
is not changed.
\end{lemma}

\begin{lemma}
$(U_t,U_x)\ne(0,0)$ for a local transformation between two equations from class~(\ref{maineq})
only if $D\in\{u^\mu,\;e^u\}\!\!\mod G^{\Equiv}.$
\end{lemma}

As an example of discrete equivalence transformations we can give the involution
\[
\tilde t=t,\ \tilde x=-x,\ \tilde u=u+\alpha x
\]
in the couple of equations
\[
f(x)u_t=(e^uu_x)_x+\alpha e^uu_x \quad \mbox{and}  \quad e^{-\alpha x}f(-x)u_t=(e^uu_x)_x+\alpha e^uu_x.
\]
Moreover, this transformation is a discrete invariance transformation for the equation
\[
g(x)e^{-\alpha x/2}u_t=(e^uu_x)_x+\alpha e^uu_x,
\]
iff $g$ is an even function.

We also investigated some transformations into other classes of reaction-diffusion equations.
So, using the discrete transformation $\tilde t=t,$ $\tilde x=-x,$ $\tilde u=u+x/2$ we can reduce the equation
\[
e^{-x/2}u_t=e^u(u_{xx}+u_x^2+u_x)
\]
to the reaction-diffusion equation from the classification of Dorodnitsyn~\cite{Dorodnitsyn}:
\[
\tilde u_t=(e^{\tilde u}\tilde u_{\tilde x})_{\tilde x}-\frac14e^{\tilde{u}}.
\]

\section{Exact solutions}\label{sec_exsol}

We now turn to presentation of some exact solutions for~(\ref{maineq}).
Using our classification with respect to all the possible local transformations
(i.e. not only with respect to ones from $G^{\Equiv}$), at first we can obtain solutions of
simpler equations (e.g. 6a from Tables~2 or~3) by means of the classical
Lie--Ovsiannikov algorithm or non-classical methods.
Then we transform them to solutions of more complicated equations (such as 6b, 6c, \dots).

Let us note that the equations with $f=1$ are well investigated and the most of exact solutions
given below were constructed before (see citations in~\cite{Polyanin&Zaitsev}).
However, at the best of our knowledge,
there exist no works containing a systematic study of all the possible Lie reductions in this
class, as well as the exhaustive consideration of integrability and exact solutions of the corresponding
reduced equations. That is why we have decided to implement the relevant Lie reduction algorithm independently,
especially since it is not a difficult problem.

So, let us consider equation~2.6a
\begin{equation}\label{solv1}
u_t=(e^uu_x)_x.
\end{equation}
Let us remind that for~(\ref{solv1}) the basis of $A^{\max}$
is formed by the operators
\[
Q_1=\p_t,\ Q_2=t\p_t-\p_u,\ Q_3=\p_x,\ Q_4=x\p_x+2\p_u.
\]
The only non-zero commutators of these operators are
$[Q_1,Q_2]=Q_1$ and $[Q_3,Q_4]=Q_3.$ Therefore $A^{\max}$ is a realization
of the algebra~$2A_{2.1}$~\cite{mubarakzyanov1963.1}.
All the possible inequivalent (with respect to inner automorphisms)
one-dimensional subalgebras of~$2A_{2.1}$~\cite{Patera} are exhausted by the ones
listed in Table~\ref{ReductionTableForEu} along with
the corresponding ansatzes and the reduced ODEs.

\begin{table} \footnotesize
\renewcommand{\arraystretch}{1.2}
\caption{Reduced ODEs for (\ref{solv1}). $\alpha\ne0,\ \varepsilon=\pm1,\ \delta=\sign t$.}
\begin{center}
\begin{tabular}{|l|l|c|c|l|}
\hline \vspacebefore
N&Subalgebra& Ansatz $u=$& $\omega$ &\hfill {Reduced ODE\hfill} \\
\hline
1&$\langle Q_3\rangle$ & $\varphi(\omega)$ & $t$ & $\varphi'=0$\\
2&$\langle Q_4\rangle$ & $\varphi(\omega)+2\ln|x|$ & $t$ & $\varphi'=2e^{\varphi}$\\
3&$\langle Q_1\rangle$ & $\varphi(\omega)$ & $x$ & $(e^{\varphi})''=0$\\
4&$\langle Q_2\rangle$ & $\varphi(\omega)-\ln|t|$ & $x$ & $(e^{\varphi})''=-\delta$\\
5&$\langle Q_1+\varepsilon Q_3\rangle$ & $\varphi(\omega)$ & $x-\varepsilon t$ &
$(e^{\varphi})''=-\varepsilon\varphi'$\\
6&$\langle Q_2+\varepsilon Q_3\rangle$ & $\varphi(\omega)-\ln|t|$ & $x-\varepsilon\ln|t|$
& $(e^{\varphi})''=-\delta(\varepsilon\varphi'+1)$\\
7&$\langle Q_1+\varepsilon Q_4\rangle$ & $\varphi(\omega)+2\varepsilon t$ & $xe^{-\varepsilon t}$ &
$(e^{\varphi})''=-\varepsilon\omega\varphi'+2\varepsilon$\\
8&$\langle Q_2+\alpha Q_4\rangle$ & $\varphi(\omega)+(2\alpha-1)\ln|t|$ & $x|t|^{-\alpha}$ &
     $(e^{\varphi})''=\delta(-\alpha\omega\varphi'+2\alpha-1)$\\
\hline
\end{tabular}
\end{center}
\label{ReductionTableForEu}
\end{table}
We succeeded in solving the equations \ref{ReductionTableForEu}.1--\ref{ReductionTableForEu}.5.
Thus we have the following solutions of~(\ref{solv1}):
\[
u=\ln|c_1x+c_0|, \qquad
u=\ln\left(\dfrac{-x^2}{2t}+\dfrac{c_1x+c_0}t\right),
\]
\[
u=\varphi(x-\varepsilon t)\quad \mbox{where}\
\int{\frac{e^{\varphi}}{c_1-\varepsilon\varphi}}d\varphi=\omega+c_0.
\]
Using them we can  construct solutions for Cases 2.6b--2.6d easily.
For example, the transformation~2.4
yields the corresponding solutions for the more complicated and interesting equation
\[
\dfrac{e^x}{(\gamma e^x+1)^3}\,u_t=(e^uu_x)_x+e^uu_x
\]
having the localized density (Case 2.6c):
\[
u=\ln\left|c_1+c_0(e^{-x}+\gamma)\right|, \quad
u=\ln\left(-\frac1{2t(e^{-x}+\gamma)}-\frac{c_1}t+c_0\frac{e^{-x}+\gamma}t\right).
\]

A singular value of the parameter~$\mu$ for Case~3.6a is $\mu=-1.$ So, the equation
\begin{equation} \label{solv3}
u_t=\left(\frac{u_x}u\right)_x
\end{equation}
is distinguished by the reduction procedure. It is remarkable that Cases~3.6c and~3.6d
are reduced exactly to the equation~(\ref{solv3}).
The invariance algebra of~(\ref{solv3}) is generated by the operators
\[
Q_1=\p_t,\ Q_2=t\p_t+u\p_u,\ Q_3=\p_x,\ Q_4=x\p_x-2u\p_u
\]
and is a realization of the algebra~$2A_{2.1}$ too. The reduced ODEs for
are listed in Table~\ref{ReductionTableFormu-1}.
\begin{table} \footnotesize
\renewcommand{\arraystretch}{1.2}
\caption{Reduced ODEs for (\ref{solv3}). $\alpha\ne0,\ \varepsilon=\pm1$.}
\begin{center}
\begin{tabular}{|l|l|c|c|l|}
\hline \vspacebefore
N&Subalgebra& Ansatz $u=$& $\omega$ &\hfill {Reduced ODE\hfill} \\
\hline
1&$\langle Q_3\rangle$ & $\varphi(\omega)$ & $t$ & $\varphi'=0$\\
2&$\langle Q_4\rangle$ & $\varphi(\omega)x^{-2}$ & $t$ & $\varphi'=2$\\
3&$\langle Q_1\rangle$ & $\varphi(\omega)$ & $x$ & $(\varphi^{-1}\varphi')'=0$\\
4&$\langle Q_2\rangle$ & $\varphi(\omega)t$ & $x$ & $(\varphi^{-1}\varphi')'=\varphi$\\
5&$\langle Q_1+\varepsilon Q_3\rangle$ & $\varphi(\omega)$ & $x-\varepsilon t$ &
      $(\varphi^{-1}\varphi')'=-\varepsilon\varphi'$\\
6&$\langle Q_2+\varepsilon Q_3\rangle$ & $\varphi(\omega)t$ & $x-\varepsilon\ln|t|$
       & $(\varphi^{-1}\varphi')'=-\varepsilon\varphi'+\varphi$\\
7&$\langle Q_1+\varepsilon Q_4\rangle$ & $\varphi(\omega)e^{-2\varepsilon t}$ & $xe^{-\varepsilon t}$ &
       $(\varphi^{-1}\varphi')'=-\varepsilon\omega\varphi'-2\varepsilon\varphi$\\
8&$\langle Q_2+\alpha Q_4\rangle$ & $\varphi(\omega)t|t|^{-2\alpha}$ & $ x|t|^{-\alpha}$ &
    $(\varphi^{-1}\varphi')'=-\alpha\omega\varphi'+(1-2\alpha)\varphi$\\
\hline
\end{tabular}
\end{center}
\label{ReductionTableFormu-1}
\end{table}
After integrating Cases~\ref{ReductionTableFormu-1}.1--\ref{ReductionTableFormu-1}.4 we obtain the following solutions
of~(\ref{solv3}):
\[
u=c_0e^{c_1x},\qquad
u=\dfrac{2c_1^2t}{\cos^2c_1(x+c_0)},\qquad
u=\dfrac{2tc_0c_1^2e^{c_1x}}{(1-c_0e^{c_1x})^2},
\]
\[
u=\dfrac{c_1}{-\varepsilon+c_0e^{c_1(x-\varepsilon t)}},\qquad
u=\frac\varepsilon{x-\varepsilon t+c_0}, \qquad
u=\frac{2t}{(x+c_1)^2+c_0t^2}.
\]
Analogously to previous case
we obtain by means of transformations~3.5 exact solutions of equation~3.6d in the following forms:
\[
u=c_0e^{(c_1-\gamma)e^{-x}},\qquad
u=\dfrac{2c_1^2te^{-\gamma e^{-x}}}{\cos^2c_1(e^{-x}+c_0)},\qquad
u=\dfrac{2tc_0c_1^2e^{(c_1-\gamma)e^{-x}}}{(1-c_0e^{c_1e^{-x}})^2},
\]
\[
u=\dfrac{c_1e^{-\gamma e^{-x}}}{-\varepsilon+c_0e^{c_1(e^{-x}-\varepsilon t)}},\qquad
u=\frac{\varepsilon e^{-\gamma e^{-x}}}
{e^{-x}-\varepsilon t+c_0}, \qquad
u=\frac{2te^{-\gamma e^{-x}}}{(e^{-x}+c_1)^2+c_0t^2}.
\]
Another example of equation with localized density is given by Case~3.6f.
To look for exact solutions for it, at first we reduce it to the equation~3.6a
\begin{equation} \label{solv5}
u_t=\left(u^{\mu}u_x\right)_x.
\end{equation}
As in the previous cases the invariance algebra of~(\ref{solv5})
\[
A^{\max}=\langle Q_1=\p_t,\ Q_2=t\p_t-\mu^{-1}u\p_u,\ Q_3=\p_x,\ Q_4=x\p_x+2\mu^{-1}u\p_u\rangle
\]
is a realization of the algebra~$2A_{2.1}$. The result of reduction~(\ref{solv5})
under inequivalent subalgebras of~$A^{\max}$ is written down in Table~\ref{ReductionTableFormuforall}.

\begin{table} \footnotesize
\renewcommand{\arraystretch}{1.2}
\caption{Reduced ODEs for (\ref{solv5}). $\mu\ne0,-1,$ $\alpha\ne0,$ $\varepsilon=\pm1,$ $\delta=\sign t.$}
\begin{center}
\begin{tabular}{|l|l|c|c|l|}
\hline \vspacebefore
N&Subalgebra& Ansatz $u=$& $\omega$ &\hfill {Reduced ODE\hfill} \\
\hline
1&$\langle Q_3\rangle$ & $\varphi(\omega)$ & $t$ & $\varphi'=0$\\
2&$\langle Q_4\rangle$ & $\varphi(\omega)|x|^{2/\mu}$ & $t$ &
   $\varphi'=2\mu^{-2}(2+\mu)\varphi^{\mu+1}$\\
3&$\langle Q_1\rangle$ & $\varphi(\omega)$ & $x$ & $(\varphi^{\mu}\varphi')'=0$\\
4&$\langle Q_2\rangle$ & $\varphi(\omega)|t|^{-1/\mu}$ & $x$
  & $(\varphi^{\mu}\varphi')'=-\delta\mu^{-1}\varphi$\\
5&$\langle Q_1+\varepsilon Q_3\rangle$ & $\varphi(\omega)$ & $x-\varepsilon t$ &
     $(\varphi^{\mu}\varphi')'=-\varepsilon\varphi'$\\
6&$\langle Q_2+\varepsilon Q_3\rangle$ & $\varphi(\omega)|t|^{-1/\mu}$ & $ x-\varepsilon\ln|t|$
  & $(\varphi^{\mu}\varphi')'=-\delta\varepsilon\varphi'-\delta\mu^{-1}\varphi$\\
7&$\langle Q_1+\varepsilon Q_4\rangle$ & $\varphi(\omega)e^{2\varepsilon\mu^{-1} t}$
       & $xe^{-\varepsilon t}$
  &$(\varphi^{\mu}\varphi')'=-\varepsilon\omega\varphi'+2\mu^{-1}\varepsilon\varphi$\\
8&$\langle Q_2+\alpha Q_4\rangle$ & $\varphi(\omega)|t|^{(2\alpha-1)/\mu}$ & $ x|t|^{-\alpha}$ &
   $(\varphi^{\mu}\varphi')'=\delta\mu^{-1}(2\alpha-1)\varphi-\delta\alpha\omega\varphi'$\\
\hline
\end{tabular}
\end{center}
\label{ReductionTableFormuforall}
\end{table}

For some of the reduced equations we can construct the general solutions.
For other ones we succeeded to find only particular solutions.
These solutions are following:
\[
u=|c_1x+c_0|^{\frac1{\mu+1}},
\quad
u=(c_0-\varepsilon\mu(x-\varepsilon t))^{\frac1\mu},
\quad
u=\left(-\frac\mu{\mu+2}\,\frac{(x+c_0)^2}{2t}+c_1|t|^{-\frac{\mu}{\mu+2}}\right)^{\frac1\mu},
\]
\[
u=\left(-\frac\mu{\mu+2}\,\frac{(x+c_0)^2}{2t}+
c_1(x+c_0)^{\frac\mu{\mu+1}}|t|^{-\frac{\mu(2\mu+3)}{2(\mu+1)^2}}\right)^{\frac1\mu},
\]

All the results of Tables~\ref{ReductionTableFormu-1},~\ref{ReductionTableFormuforall} as well as
constructed solutions can be extended to equations~3.6b--3.6g using the local
equivalence transformations.
So for the equation
\begin{equation} \label{solv6}
\dfrac{e^{-2x}}{(e^{-x}+\gamma)^{\frac{4+3\mu}{1+\mu}}}\,u_t=\left(u^{\mu}u_x\right)_x+u^{\mu}u_x
\end{equation}
(Case~3.6f) the transformations 3.7 yield exact solutions in the form
\[
u=|c_0(e^{-x}+\gamma)-c_1|^{\frac1{\mu+1}},
\quad
u=\left(c_0+\frac{\varepsilon\mu}{e^{-x}+\gamma}+\varepsilon^2\mu t\right)^{\frac1\mu}
|e^{-x}+\gamma|^{-\frac1{\mu+1}},
\]
\[
u=\left(-\frac\mu{\mu+2}\,\frac1{2t}\left(c_0-\frac1{e^{-x}+\gamma}\right)^2
+c_1|t|^{-\frac{\mu}{\mu+2}}\right)^{\frac1\mu}
|e^{-x}+\gamma|^{-\frac1{\mu+1}},
\]
\[
u=\left(-\frac\mu{\mu+2}\,\frac1{2t}\left(c_0-\frac1{e^{-x}+\gamma}\right)^2
+c_1\left(c_0-\frac1{e^{-x}+\gamma}\right)^{\frac\mu{\mu+1}}
|t|^{-\frac{\mu(2\mu+3)}{2(\mu+1)^2}}
\right)^{\frac1\mu}|e^{-x}+\gamma|^{-\frac1{\mu+1}},
\]

A number of exact solutions were constructed for equations from class~(\ref{maineq1}) ($f=1$)
by means of nonclassical methods. Starting from them and using local transformations of
conditional equivalence, we can obtain non-Lie exact solutions for more complicate
equations (Cases 6b, 6c,~\dots).

So, N.K.~Amerov~\cite{Amerov} and J.R.~King~\cite{King1} suggested to look for solutions of the equation
$u_t=(u^{-1/2}u_x)_x$
(3.6a, $\mu=-1/2$) in the form $u=(\varphi^1(x)t+\varphi^0(x))^2$ where the functions $\varphi^1(x)$ and $\varphi^0(x)$
satisfy the system of ODEs
$\varphi^1_{xx}=(\varphi^1)^2,$ $\varphi^0_{xx}=\varphi^0\varphi^1.$
A particular solution of this system is
\[
\varphi^1=\frac6{x^2} , \qquad  \varphi^0=\frac{c_1}{x^2}+\frac{c_2}{x^3}.
\]
And the corresponding solution for the equation~3.6f with $\mu=-1/2$ can be written down as
\[
u=(6t+c_1'+c_2e^{-x})^2(e^{-x}+\gamma)^6.
\]

\section{Conclusion}
In this paper group classification in the class of equations~(\ref{maineq}) is performed completely.
The main results on classification are collected in Tables~\ref{dall}--\ref{dum} where we list
inequivalent cases of extensions with the corresponding Lie invariance algebras.
Among the presented equations there exist ones which have the density $f$ localized
in the space of $x$ and are invariant with respect to more abundant Lie algebras than $A^{\ker}$.
After the tables we write down all the additional equivalence transformations
reducing some equations from our classification to other ones of simpler forms.
(In fact, the equations of~(\ref{maineq}) have been classified with respect to two different equivalence
relations generated by either the equivalence group or the set of all possible transformations.)
For a number of equations from the list of the reduced ones we construct optimal systems of
inequivalent subalgebras, corresponding Lie ansatzes and exact invariant solutions.
By means of additional equivalence transformations the obtained solutions are transformed
to the ones for the more interesting and complicated equations with localized densities.

We describe equivalence transformations within a class of PDEs~(\ref{maineq}) using
the infinitesimal and direct methods.
The direct method enables us to find the group of all the possible local equivalence
transformations (i.e. not only continuous ones)  in the whole class~(\ref{maineq})
as well as all the conditional equivalence transformations.
Moreover, we begin to solve the general equivalence problem for any pair of equations
from class~(\ref{maineq}) with respect to the local transformations (Lemmas~1--4).

We plan to continue investigations of this subject. For the class under consideration we plan to
perform classification of potential and nonclassical (conditional) symmetries
and finish studying all the possible partial equivalence transformations.
We also intend to investigate
existence, localization and asymptotic properties of solutions of initial and boundary-value
problems for nonlinear convection-diffusion equations.

\medskip

{\bf Acknowledgments.} The authors are grateful to
Profs. V.~Boyko, A.~Nikitin, A.~Sergyeyev,  I.~Yehorchenko and A.~Zhalij
for useful discussions and interesting comments.

The research of NMI was partially supported by National Academy of Science of Ukraine
in the form of the grant for young scientists.

\end{document}